\documentclass[prd,twocolumn,nofootinbib,showpacs,superscriptaddress]{revtex4}

\usepackage{amsfonts}
\usepackage{amsmath}
\usepackage{bm}
\usepackage{graphicx}
\usepackage[latin1]{inputenc}
\usepackage{latexsym}
\usepackage{rotating}
\usepackage{dcolumn}      
\usepackage{longtable}    

\newcommand{\be}{\begin{equation}}
\newcommand{\ee}{\end{equation}}
\newcommand{\bea}{\begin{eqnarray}}
\newcommand{\eea}{\end{eqnarray}}
\def\bbh#1{binary black hole#1 (BBH#1)\gdef\bbh{BBH}}
\def\bh#1{black hole#1 (BH#1)\gdef\bh{black hole}}
\def\qnm#1{quasinormal mode#1 (QNM#1)\gdef\qnm{quasinormal mode}}

\begin{document}

\title{Probing the Binary Black Hole Merger Regime with Scalar Perturbations}

\author{Eloisa Bentivegna}
\author{Deirdre M. Shoemaker}
\author{Ian Hinder}
\author{Frank Herrmann}
\affiliation{Center for Gravitational Wave Physics, \\
Institute for Gravitation and the Cosmos and Department of Physics,\\
The Pennsylvania State University, University Park, PA 16802, USA}

\date{\today}

\begin{abstract}
We present results obtained by scattering a scalar
field off the curved background of a coalescing binary black hole system. 
A massless scalar field is evolved 
on a set of fixed backgrounds, each provided by a spatial hypersurface generated 
numerically during a binary black hole merger. We show that the 
scalar field scattered from the merger region exhibits quasinormal ringing once a common 
apparent horizon surrounds the two black holes. This occurs earlier
than the onset of the perturbative regime as measured by the start of the quasinormal ringing in
the gravitational waveforms.
We also use the scalar quasinormal frequencies to associate a mass and a spin 
with each hypersurface, and observe the compatibility of this measure with 
the horizon mass and spin computed from the dynamical horizon framework. 
\end{abstract}


\preprint{IGPG-07}

\maketitle

\section{Introduction}\label{intro}

Over the past thirty years, numerical evolutions of \bbh{} systems have
provided deeper and deeper insight into the properties of the vacuum two-body 
problem of general relativity. Today a number of 
codes~\cite{Pretorius:2005gq,Campanelli:2005dd,Baker:2005vv,Herrmann:2007zz,Gonzalez:2006md,Koppitz:2007ev,Pfeiffer:2007yz,Tichy:2007hk}
are able to accurately evolve black hole binaries; in particular, the merger regime itself is now available for studies.
It was shown in reference~\cite{BCP,multipolar} that the merged system seems to enter the perturbative 
stage while it is still evolving and radiating energy and angular momentum.
This is in agreement with earlier results from the Close Limit approximation~\cite{Price:1994pm} and 
from the Lazarus project~\cite{Baker:2001sf}.

In order to study the transition between the nonlinear merger stage and the final
ringing black hole, we scatter a test scalar field off the geometry generated in the inspiral and merger of the binary system.
Test fields and perturbation theory have traditionally been used to study excitations
on a single stationary black hole spacetime, both for the Schwarzschild~\cite{Regge:1957td,Zerilli:1970se,Vishveshwara:1970}
and for the Kerr solutions~\cite{teukolsky}.
Numerical methods were used in several 
scenarios (see for instance~\cite{Krivan:1996da} and~\cite{Dorband:2006gg}); more recently, these results have been extended to 
more general black hole solutions, including non-vacuum spacetimes~\cite{Shao:2004ws} and models with 
non-flat asymptotic topology~\cite{Molina:2003dc,Chan:1996yk}.

An interesting result of \qnm{} evolution on {\it non-stationary}, spherically 
symmetric black hole 
spacetimes is that the mass extracted from the complex quasinormal frequency tracks the time-dependent \bh{}
mass quite closely (modulo a delay effect),
equaling the constant quasinormal frequency one would obtain on a static 
spacetime with the mass parameter equal to the instantaneous mass at each time~\cite{Shao:2004ws}.
This adiabatic picture has an intriguing parallel in \bh{} thermodynamics:
the application of the first law to a spacetime containing an axisymmetric 
dynamical horizon shows that the mass increase between two cross sections 
with horizon area $A_1$ and $A_2$ and angular momentum
$J_1$ and $J_2$ is given exactly by $M(A_2, J_2)-M(A_1, J_1)$, where 
$M(A,J)$ represents the functional dependence of the mass $M$ on $A$ and $J$
for a Kerr solution~\cite{ih}.
In other words, as far as the first law of thermodynamics is concerned, 
this portion of the spacetime can formally be pictured as a stack of spatial 
hypersurfaces, with each slice belonging to a member of the Kerr family with a 
different mass and spin parameter.

In this study we aim to address some questions concerning
the transition between a system of two inspiraling black holes and a
single perturbed black hole.  We investigate how early during the \bbh{} evolution the scalar
field probe exhibits the quasinormal ringing phase ordinarily expected
from a spacetime containing a single black hole, and what features this
phase displays when compared to gravitational wave quasinormal
ringing of the single stationary black hole that is left behind by the
coalescence process.
Based on these considerations we present a novel strategy that
consists of (i) simulating a \bbh{} merger and extracting horizon
information from the merged \bh{}, (ii) taking snapshots of the
spatial geometry obtained in the \bbh{} merger at a sequence of
coordinate times, (iii) evolving a massless scalar field on the
corresponding spatial hypersurfaces, and (iv) observing the scalar
field undergo a quasinormal ringing phase and extracting its complex
frequency, which in turn can be converted into a mass and a spin
parameter using the functional dependence of the quasinormal 
frequencies on these two parameters prescribed by perturbation theory.
We then compare these estimates to the mass and spin
extracted directly from the horizon.

This method has the benefit of effectively separating the evolution
time of the binary system from the relaxation time of the scalar
perturbations, allowing for a measure of the ``instantaneous''
properties of the foliation.  Notice that this procedure is effectively
equivalent to the evolution of a scalar field on the fictitious
spacetime one would obtain by selecting each spatial surface and
freezing it in time. This of course is not meant to constitute a 
physical solution to the coupled scalar-gravitational system:
as emphasized above the scalar field is used
here exclusively as a probing agent for individual hypersurfaces.

We will present a brief overview of scalar perturbations in
section \ref{sec:scalar}, along with the method we use
to evolve the scalar wave equation on a numerically-generated curved
background, extract the complex frequencies and associate a mass and 
spin parameter to them.  Section \ref{sec:results} contains the results of 
our scalar field evolutions, along with the corresponding waveforms and
analysis of errors. Finally, in our concluding section \ref{Conclusions}, we
will discuss the results.


\section{Scalar field evolution}\label{sec:scalar}
The behavior of a massless scalar field on the background
of a single Kerr black hole has been the subject of extensive
analytical treatments. A good introduction is given in
reference~\cite{Chandrasekhar:1985kt} and a thorough treatment of the subject can be
found in Nollert's~\cite{Nollert:1999ji} and Kokkotas and Schmidt's review works~\cite{Kokkotas:1999bd}. 

\subsection{Quasinormal ringing}
The behavior of a massless scalar field $\Phi$ on the 
background provided by a Kerr black hole of mass $M$ and Kerr spin parameter $a=J/M$
is governed by the source-free scalar wave equation:
\be
\label{eq:KG}
\square \Phi := g_{ab} \nabla^a \nabla^b \Phi = 0 \,.
\ee
Separating out the angular variables, one reaches a wave equation with a potential~\cite{teukolsky}.
Upon further study, one finds that the field exhibits many of the well-known features of waves impinging on 
a potential; for instance, it exhibits quasinormal ringing, i.e. a phase of decaying 
oscillations characterized by a certain complex frequency spectrum.

The complex nature of the frequency spectrum leads to several undesirable 
consequences, some of which have direct ramifications on our ability to detect and 
characterize quasinormal ringing in a numerical simulation:
first and foremost, the spectrum of quasinormal modes does not constitute a 
complete set of eigenfunctions~\cite{Nollert:1999ji,Ching:1995rt,Chandrasekhar:1985kt}: 
early on, the behavior of the scalar field depends 
on the process that generated the perturbation, and the field is said to undergo the 
{\it prompt response} regime, where $\Phi$ still displays memory of the perturbing
mechanism~\cite{Ching:1999fb}. Additionally, at late times the backscattering off the curvature will 
overtake the by-now feeble damped oscillations and dominate the waveforms, generating 
the so-called {\it tails}~\cite{Ching:1999fb}. A Green's-function analysis of the origin
of these three phases, along with a convergence study of the quasinormal mode expansion is presented 
in~\cite{Leaver:1986gd} for Schwarzschild and in~\cite{Berti:2006wq} for Kerr black holes. 
The question then arises about how to identify the onset time of quasinormal ringing, since 
an accurate determination of this quantity is of the utmost importance while fitting 
for the quasinormal frequencies. This issue is usually referred to as the {\it time shift} 
problem; for a discussion of its influence on the bias of the fitted parameters, see~\cite{Dorband:2006gg}.
Also, concepts such as the fraction of a given waveform that can be identified as each quasinormal mode
and the gravitational energy contained in each quasinormal mode cannot be defined in rigorous 
terms~\cite{Nollert:1998ys}; the absence of such auxiliary notions also poses serious 
complications to identifying the onset of quasinormal ringing in a numerical waveform, and 
quantifying its quasinormal content. Berti et al.~\cite{multipolar} review this issue from
the numerical point of view.

Notice that, due to their dependence on the mass $M$ and spin parameter $j=a/M$ 
of the background Kerr solution, the quasinormal ringing portion of the scattered 
waves carries information about $M$ and $j$ away from the scattering center
(a property which, for instance, makes gravitational quasinormal ringing a powerful tool 
for gravitational wave astronomy~\cite{Berti:2005ys}). 

In this work, we use scalar perturbations to probe 
the parameters describing the final black hole that results from the binary coalescence. 
In order to provide as accurate a depiction of the black hole as possible, we will first 
have to devise a strategy to tackle the ambiguities associated with quasinormal modes and find 
a reliable strategy to detect and measure the quasinormal components in a waveform. 
In section \ref{sec:qnm} we will present our experimental method.

\subsection{Scalar field evolution on a numerically generated background}\label{sec:sim}
We evolve a scalar field on a {\it numerically generated} background
obtained from a binary black hole evolution. We take snapshots of the
spatial geometry and then use each of these as a time-independent background
on which the scalar field propagates.

During the course of the evolution, we store the values of the grid functions 
corresponding to the gravitational variables at a sequence of coordinate times 
$\left\{t_N\right\}$, which we denote with a bar:
\bea
\bar\gamma_{ij}(x^k) \equiv \gamma_{ij}(t_N,x^k)\\
\bar K_{ij}(x^k) \equiv K_{ij}(t_N,x^k)\\
\bar\alpha(x^k) \equiv \alpha(t_N,x^k)\\
\bar\beta^i(x^k) \equiv \beta^i(t_N,x^k)
\eea
and evolve a scalar field on the background provided by the corresponding
{\it frozen} geometry. In this sense, $\Phi$ is nothing more than a probing agent 
for each of the spatial slices on which we perform the experiment, and is meant 
solely to provide information, through its scattering, about the curvature of 
each single slice rather than to explore a scalar-gravitational coupling 
of an actual physical system.

\subsection{Background spacetime}

The evolution of the gravitational variables $\gamma_{ij}$ and $K_{ij}$
is performed according to the BSSN system~\cite{Nakamura,Shibata:1995we,Baumgarte:1998te}, 
as implemented in the PSU code through the Kranc generator~\cite{Husa:2004ip}. The fourth order finite 
differencing code is based on the Cactus infrastructure~\cite{cactus} 
and is endowed with Carpet's mesh refinement capabilities~\cite{carpet}.
The gauge variables $\alpha$ and $\beta^i$ are initialized and evolved according 
to the moving puncture prescription~\cite{Campanelli:2005dd,Baker:2005vv}, while 
the initial data are generated following~\cite{Ansorg:2004ds}.

For the binary coalescence runs, we choose an initial configuration 
usually referred to as R1~\cite{Baker:2006yw}: 
this setup represents two 
non-spinning black holes with irreducible masses equal to $0.505 \mathcal{M}$
(where $\mathcal{M}$ denotes a mass scale related to the ADM mass via $E_{\mathrm{ADM}}/\mathcal{M}=0.9957$).
The two \bh{s} start at a coordinate 
separation of $6.514\mathcal{M}$, with linear 
momenta perpendicular to the separation vector and equal to $0.133 \mathcal{M}$.
Further details regarding our code and our earlier study of this system can be found in~\cite{Vaishnav:2007nm}.
We used a reference spatial resolution of $\mathcal{M}/51.2$, 
and evolved some simulations at two extra resolutions
($\mathcal{M}/44.8$ and $\mathcal{M}/57.6$) to obtain an estimate for the 
truncation errors (see section \ref{sec:errors} below).

Figure \ref{fig:psi4} plots the gravitational wave from 
the simulated system: the two black holes inspiral around each other for about two orbits 
before plunging and forming a single final perturbed black hole whose parameters 
can be inferred, due to energy and angular momentum conservation, from the emitted 
radiation in terms of the Newman-Penrose scalar $\Psi_4$, as discussed in~\cite{Campanelli:1998jv}.
For the $\mathcal{M}/51.2$ resolution run, starting with an initial ADM mass and angular
momentum equal to $E_{\mathrm{ADM}}/\mathcal{M}=0.9957$ and $J_{\mathrm{ADM}}/\mathcal{M}^2=0.862 $,
the system radiates $E_{\mathrm{rad}}/\mathcal{M}=0.032 \pm 0.005$ and $J_{\mathrm{rad}}/\mathcal{M}^2=0.22 \pm 0.01$,
leaving behind a final black hole with $E_f/\mathcal{M}=0.963 \pm 0.005$ and
$J_f/\mathcal{M}^2=0.64 \pm 0.01$,
in agreement with the estimates of mass and spin 
provided by the analysis of the fundamental quasinormal tone of $\Psi_4^{22}$
($E_{\mathrm{QNM}}/\mathcal{M}=0.957 \pm 0.005$ and $J_{\mathrm{QNM}}/\mathcal{M}^2=0.62 \pm 0.01$)
and with previous results obtained with other codes~\cite{Baker:2006yw}.

\begin{center}
\begin{figure}
\includegraphics[width=0.5\textwidth]{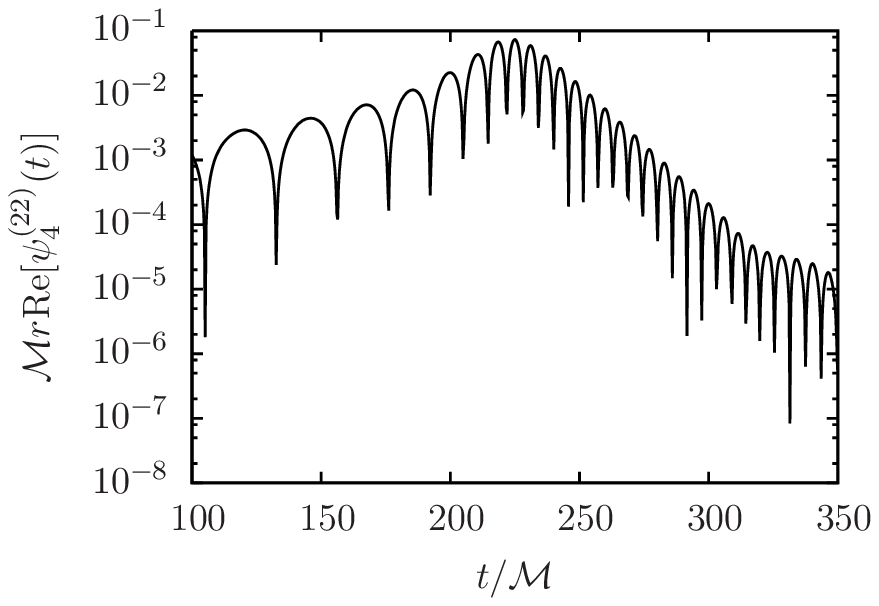}
\caption{The real part of the $\ell=2$, $m=2$ mode of the
  Newman-Penrose scalar $\Psi_4$, extracted at $r=50\mathcal{M}$. Note
  the quasinormal ringdown exhibited by the linear decay in the
  amplitude and the constant frequency on this log-linear
  plot.\label{fig:psi4}}
\end{figure}
\end{center}

\subsection{Scalar evolution in 3+1 form}

The scalar field code was developed using Kranc~\cite{Husa:2004ip} to generate the
initial data and evolution modules and the interface with the BSSN code.
A 3+1 decomposition of the scalar wave evolution equation (\ref{eq:KG}) 
in terms of $\bar\gamma_{ij}$, $\bar K^{ij}$, $\bar\alpha$ and $\bar\beta^i$
reads:
\bea
(\partial_\tau-\bar\beta^k \partial_k)\Phi&=&\bar\alpha \Psi\\
(\partial_\tau-\bar\beta^k \partial_k)\Psi&=&\frac{\bar\alpha}{\sqrt{\bar\gamma}}
\partial_i(\sqrt{\bar\gamma} \bar\gamma^{ij}\partial_j \Phi)\nonumber\\
&+&\bar\alpha \bar K \Psi + \bar\gamma^{ij} \partial_i \bar\alpha \partial_j \Phi
\eea
where we have introduced the auxiliary variable $\Psi$ in order to reduce the 
system to a first-order-in-time form. Notice that we label the time coordinate in the 
above two equations as $\tau$, in order to stress the distinction between this and
the time coordinate $t$ associated with the \bbh{} evolution.

\begin{figure}
\includegraphics[width=0.5\textwidth]{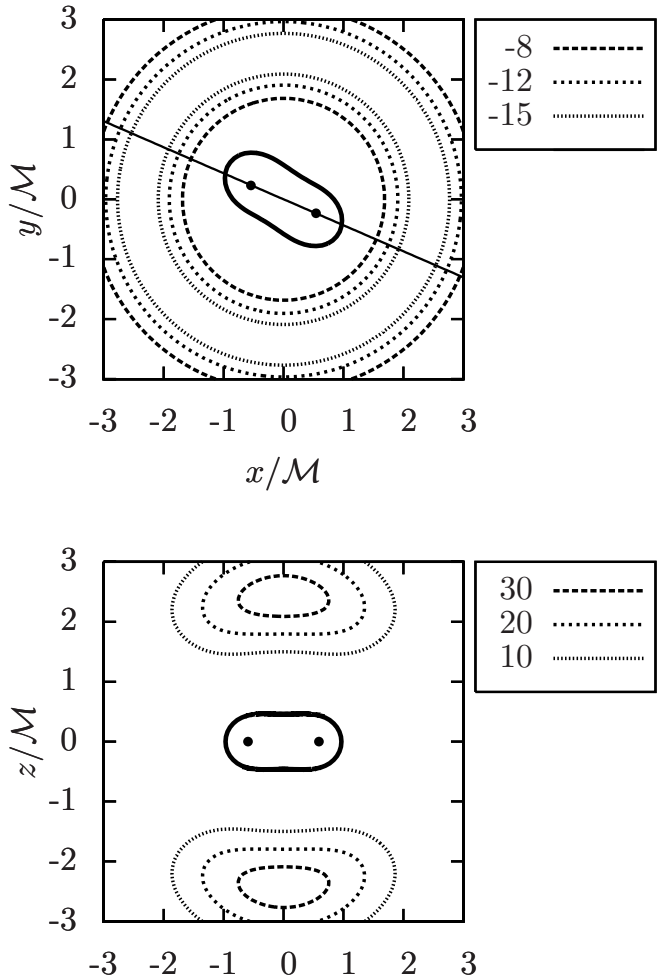}
\caption{Initial isosurfaces for the scalar field $\Phi$ as in equation 
(\ref{eq:phiid}), in the $\ell=2$ case (broken lines). The solid outline
represents the apparent horizon's intersection with the $xy$-plane
(top) and its intersection with the plane containing the $z$-axis and the 
line connecting the two black hole centers (bottom), at $t=160\mathcal{M}$. Given the initial data for $\Psi$ in equation 
(\ref{eq:psiid}), the 
two lobes along the z-axis start moving towards the origin. The detector sphere used for extraction is located at $r=5$.
\label{fig:id}}
\end{figure}

For each of the scalar evolutions, we provide initial data for $\Phi$ and $\Psi$ 
as follows:
\bea
\Phi &=& r^4 e^{-\frac{(r-r_0)^2}{\sigma^2}} \textrm{Re}[Y_{\ell 0}]\label{eq:phiid}\\
\Psi &=& -\frac{2(r-r_0)}{\alpha \sigma^2} \Phi \label{eq:psiid} 
\eea
where $r_0=1.6\mathcal{M}$ and $\sigma=1\mathcal{M}$.
An outline of our initial setup is provided in figure \ref{fig:id}.

Notice that, as observed in~\cite{Dorband:2006gg}, the fact that 
$\textrm{Re}[Y_{\ell m}]=(Y_{\ell m}+Y_{\ell,-m})/2$ implies the concurrent presence of the 
($\ell$, $m$) and ($\ell$, $-m$) modes in the initial data.
In order to produce ringing waveforms with a single dominant frequency, we chose 
only $m=0$ modes.
Following Dorband et al.~\cite{Dorband:2006gg},
we extract the multipoles of the scalar field at a coordinate radius 
$r=5\mathcal{M}$ in all cases (the effect of the extraction radius on the 
quasinormal frequency is illustrated in section \ref{sec:errors}).

\subsection{Extracting \bh{} properties from QNM ringing}\label{sec:qnm}
In the discussion above, we have mentioned how scalar perturbation
theory on a black hole background predicts that, as soon as the
initial impulse driving the perturbation propagates away, the scalar
field will relax back to its equilibrium configuration through a
period of damped (quasinormal) oscillations; numerical simulations
(see, e.g.,~\cite{Dorband:2006gg}) confirm this picture. In the
quasinormal regime the scalar field can be quite accurately described
in terms of a set of quasinormal modes identified by the three numbers
$\ell$, $m$ (labeling the angular eigenstates) and $n$ (labeling the
radial eigenstates).  In time, each multipole evolves perturbatively,
independently of the others and is characterized by a specific complex
frequency $\Omega_{\ell m n}$, whose dependence on the black hole mass
and spin is known. In principle, then, it should be
possible to extract information about the black hole parameters $M$ and $j$
by measuring $\Omega_{\ell m n}$
for a number of different $\ell m n$-modes.

The peculiar nature of quasinormal modes complicates this
procedure: since quasinormal modes do not constitute a complete set of
eigenstates, they cannot serve as a basis for the solutions of
equation (\ref{eq:KG}), and there is no guarantee that, at any given
instant of time, the scalar field can be described as a superposition
of quasinormal modes alone. However, the practical evidence is that
the system will follow a time evolution path which, for some
transitory period, will live almost entirely on the subspace spanned
by the quasinormal modes~\cite{Leaver:1986gd, Berti:2006wq}.

As in~\cite{BCP}, we compute the time series $\Phi_{\ell 0}(\tau)$: 
\bea
\Phi_{\ell 0}(\tau) &=& \int_{S_\rho} Y^*_{\ell 0}(\theta,\phi) \Phi(\rho, \theta, \phi, \tau) d^2 \Omega
\eea
(where $Y_{\ell 0}$ are the standard (spin zero) spherical harmonics with $m=0$, and 
the angular modes $\Phi_{\ell 0}$ are extracted on a sphere $S_\rho$ of 
fixed coordinate radius $\rho$), and fit it to a superposition of quasinormal modes:
\begin{eqnarray}
\label{eq:fitfunc}
&&\Phi_{\ell 0}^\mathrm{{QN}}(\tau;A,M,j,\phi,\tau_0) = \nonumber\\
&&\textrm{Re}\left[\sum_{n} A_{\ell 0 n}e^{-i \Omega_{\ell 0 n}(M,j) (\tau-\tau_0) 
+ \phi_{\ell 0 n}}\right]
\end{eqnarray}
where $A_{\ell 0 n}$ and $\phi_{\ell 0 n}$ are real constants, and the complex 
quasinormal frequencies $\Omega_{\ell 0 n}(M,j)=\omega_{\ell 0 n}(M,j)+i\alpha_{\ell 0 n}(M,j)$ depend on $M$ and $j$ 
in a somewhat complicated way, which can however be represented quite conveniently
through the use of tabulated data~\cite{Berti:2006wq}, complemented with 
third order interpolation.  Note that we only fit to the $m=0$ fundamental mode ($n=0$), however, we study the
error associated with not including higher overtones.

In order to extract the \bh{} parameters we minimize the residual of the fit
\begin{eqnarray}
\label{eq:q}
&&q(A,M,j,\phi,\tau_0,\tau_f)=\nonumber \\
&&\frac{\int_{\tau_0}^{\tau_f}d\tau \, |\Phi_{\ell 0}(\tau)-\Phi_{\ell 0}^{\mathrm{QN}}(\tau;A,M,j,\phi,\tau_0)|}{\int_{\tau_0}^{\tau_f}d\tau \, |\Phi_{\ell 0}(\tau))|}
\end{eqnarray}
with respect to $A$, $M$, $j$ and $\phi$ (with the $M$ and $j$ dependence given by the table interpolation method mentioned above). 

As mentioned above, there is no {\em a priori} prescription for the initial and final times $\tau_0$ and $\tau_f$ delimiting 
the fitting window in which the minimum search is performed. In~\cite{Dorband:2006gg} and~\cite{BCP}, $\tau_0$ is 
included as an additional fitting parameter, along with some heuristic prescription for $\tau_f$.
In~\cite{multipolar}, the value of $\tau_0$ was obtained by cross-correlating different angular modes, 
with two different cutoff criteria for $\tau_f$.
In this work, we start by choosing the initial and final times $\tau_0$ and $\tau_f$ 
so that the fitting window excludes the initial prompt response phase and the 
final wave portion contaminated by numerical and outer boundary error, 
or $\tau_0 \sim 25 \mathcal{M}$ and $\tau_f \sim 55 \mathcal{M}$.
We then perform a two-dimensional analysis of the variations in the best-fit parameters
induced by a change in $\tau_0$ and $\tau_f$.  This analysis provides an estimate
of the frequency extraction error and ensures that small variations on either of 
these parameters have no appreciable influence on the fit results.

As noted in~\cite{Berti:2007dg}, the extraction of the complex exponentials 
from a quasinormal ringing waveform is a delicate procedure. In order to minimize equation (\ref{eq:q}), we have tested
the non-linear least-squares Levenberg-Marquardt algorithm, which proved to be inadequate
for our purposes.
We then chose a fitting algorithm based on an optimization 
method known as 
the Covariant Matrix Adaptation Evolutionary Strategy (CMA-ES), 
whose details can be found in~\cite{cmaes} and which is available as a convenient open-source Matlab routine~\cite{cmaeswww}. 
As a consistency check, we have tested the fitting procedure employing CMA-ES,
the Matlab Levenberg-Marquardt \texttt{nlinfit} routine, and obtained
comparable results whenever the latter routine converged to a minimum.
We have also tested CMA-ES on synthetic waveforms such as those discussed in~\cite{Berti:2007dg}, obtaining similar results to the Kumaresan-Tufts and matrix 
pencil methods presented there.

Notice that equation (\ref{eq:fitfunc}) involves an infinite set of overtones, i.e. different
$n$-modes.  In practice, however, the excitation of the vast majority 
of modes will be highly suppressed, and we will only have to consider 
the overtones with contributions above the level of numerical error.
Recent evidence~\cite{BCP} shows that, even though higher overtones are usually heavily 
suppressed, their inclusion in the fitting
routine noticeably improves the quality of the fit.
Since our primary interest is in extraction of the \bh{} parameters, however,
we have decided to focus
on the frequency of the fundamental mode alone, choosing 
a $\tau_0$ sufficiently large to effectively guarantee that the contribution of the overtones (which damp quickly)
is below our numerical error. We discuss the error contribution from using the fundamental mode below in section \ref{sec:errors}.

Other important error sources include the finite radial and angular resolution, the presence of numerical 
error from finite differencing, and possibly additional effects due to the dynamical slicing of the spacetime 
which ordinarily occurs during a 3+1 evolution. In order to address these potential 
issues, in section \ref{sec:errors} we try to quantify their effect. 
The mass and spin estimates from independent sources also provide evidence that 
our cumulative error budget is correctly estimated. 


\section{Results}\label{sec:results}

As a first test of the validity of our setup we evolved a single
Schwarzschild black hole in isotropic coordinates using the standard
moving puncture recipe that leads to coordinate dynamics. We then
took snapshots of the gravitational field variables and solved the
scalar wave equation numerically on these hypersurfaces and extracted
the fundamental quasinormal frequency found in the scattered scalar
field for a variety of choices of $\tau_0$ and $\tau_f$ in
Equation~(\ref{eq:fitfunc}). The values of the real and imaginary part of
the quasinormal frequency coincided with the result from perturbation
theory up to $\Delta \omega \lesssim 0.002 \mathcal{M}^{-1}$
and $\Delta \alpha \lesssim 0.002 \mathcal{M}^{-1}$ (which
corresponds to $\Delta M \lesssim 0.01\mathcal{M}$ and $\Delta j
\lesssim 0.05$ for the $\ell=2$ mode).

In the binary coalescence runs, for all $t \geq 160 \mathcal{M}$ 
a common apparent horizon is present, and becomes approximately axisymmetric at $t \sim 165 \mathcal{M}$ 
(see~\cite{Dreyer:2002mx} for a description of the Killing vector field finding algorithm we 
deployed).
Starting at this time, and until after the coalescence is complete 
(at about $t=260\mathcal{M}$, as indicated by the horizon parameters 
$M$ and $j$ becoming constant in time), we freeze the gravitational 
variables at 
regular intervals and perform an evolution of the scalar field $\Phi$ as described by equation (\ref{eq:KG}) on 
the corresponding geometry, using the reference resolution of $\mathcal{M}/51.2$.

\subsection{Waveforms}

\begin{figure*}
\includegraphics[height=0.65\textwidth]{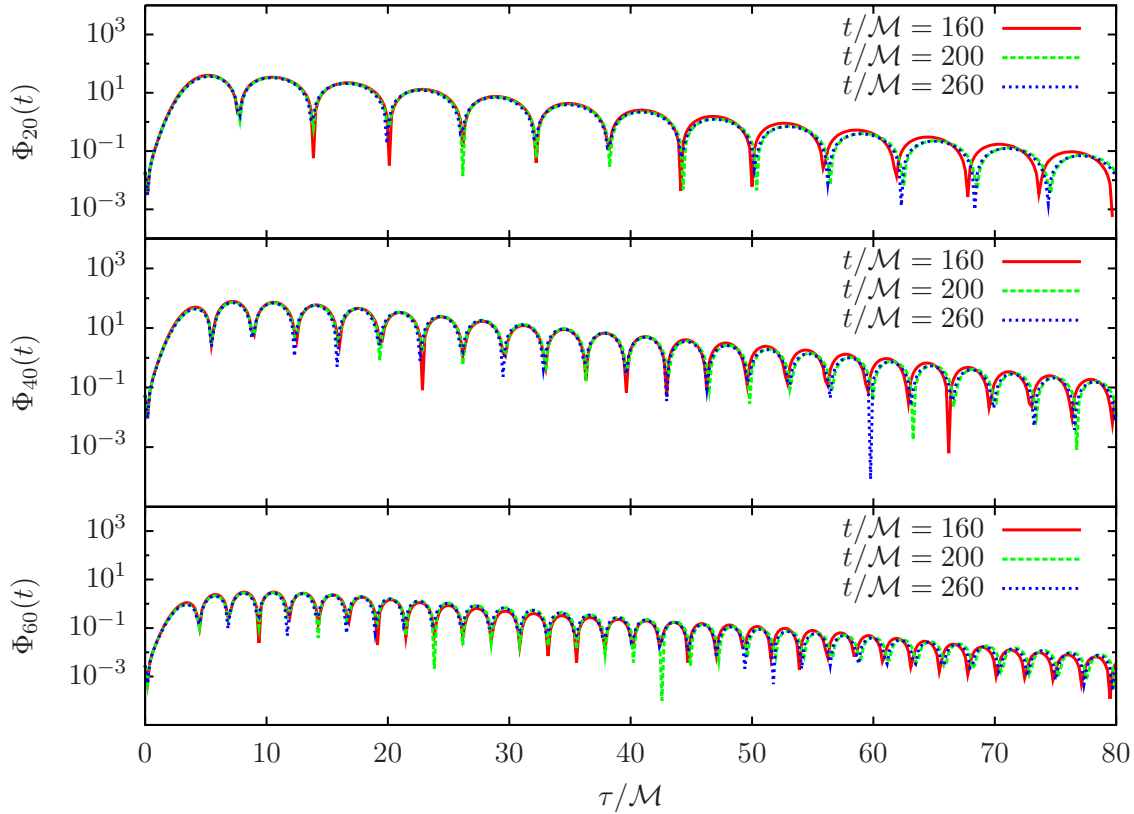}
\caption{The $\ell=\{2,4,6\}$-modes of the scalar field are shown in
  the three panels. Each mode was evolved on hypersurfaces extracted
  at $t=\{160,200,260\}\mathcal{M}$ from a binary black hole inspiral.
  The modes were obtained on a sphere of constant coordinate radius
  $\rho=5\mathcal{M}$.}
\label{fig:swave}
\end{figure*}

Figure \ref{fig:swave} shows the $\ell=\{2,4,6\}$ modes of the scalar
field extracted on a coordinate sphere of radius $\rho=5\mathcal{M}$
in separate panels. Each mode was evolved on three different
hypersurfaces labeled by $t=\{160,200,260\}\mathcal{M}$ (as stated
above, at $t=160\mathcal{M}$ the first common apparent horizon is found, and by
$t=260\mathcal{M}$ the system has settled down and we expect to
observe the typical evolution of a scalar field scattered from a Kerr
\bh{}). As can be clearly seen from this figure each waveform undergoes
a damped oscillatory phase. We observe the emergence of quasinormal
ringing on each spatial slice containing a common apparent horizon,
regardless of whether the merged system has already settled down
($t=260\mathcal{M}$) or is still evolving very dynamically
($t=160\mathcal{M}$). Initially the scalar field ringing is
virtually identical on each of the three hypersurfaces
shown, but significant dephasing is obtained towards the end of the
scalar field simulation due to the different values of the 
fundamental frequency in the three cases.  This is not an artifact of resolution, but
rather shows the difference between the hypersurfaces.

\begin{figure}
\includegraphics[width=0.48\textwidth]{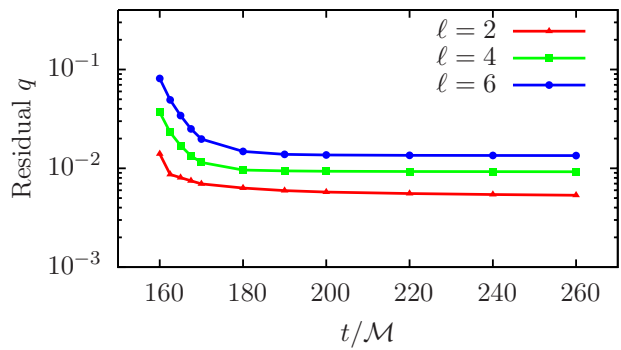}
\caption{The minimum of the fitting residual $q$ (see
  Equation~(\ref{eq:q})) as a function of $t$ for the
  $\ell=\{2,4,6\}$ mode of the scalar field. At early times $t$ the \bh{}
  has not yet settled down and hence the fit residual $q$ is larger as
  the scalar field evolved on the \bh{} background does not exhibit
  similarly clean ringdown as it does for later times.}
\label{fig:fit}
\end{figure}

For each time $t$ where we extract a hypersurface snapshot and evolve
the scalar field, we fit a quasinormal mode to the scattered scalar
field. Figure \ref{fig:fit} shows the fit residual $q$ as a function
of $t$ for the $\ell=\{2,4,6\}$ modes. This gives us a rough indicator of
how well the scattering can be described by quasinormal mode
ringing. The fit residual $q$ decreases with increasing $t$,
indicating that the evolution on earlier hypersurfaces leads to an
inferior fitting quality. This loss in fit quality at early times is
reasonable because the scalar field is expected to depart from the
familiar quasinormal ringing behavior since earlier hypersurfaces are
representing a \bh{} in more and more dynamical phases. Also note that
the higher modes are more difficult to resolve and hence a larger
residual is observed.

\subsection{Black hole parameters extracted from the scalar field probe}
\begin{figure*}
\includegraphics[width=0.8\textwidth]{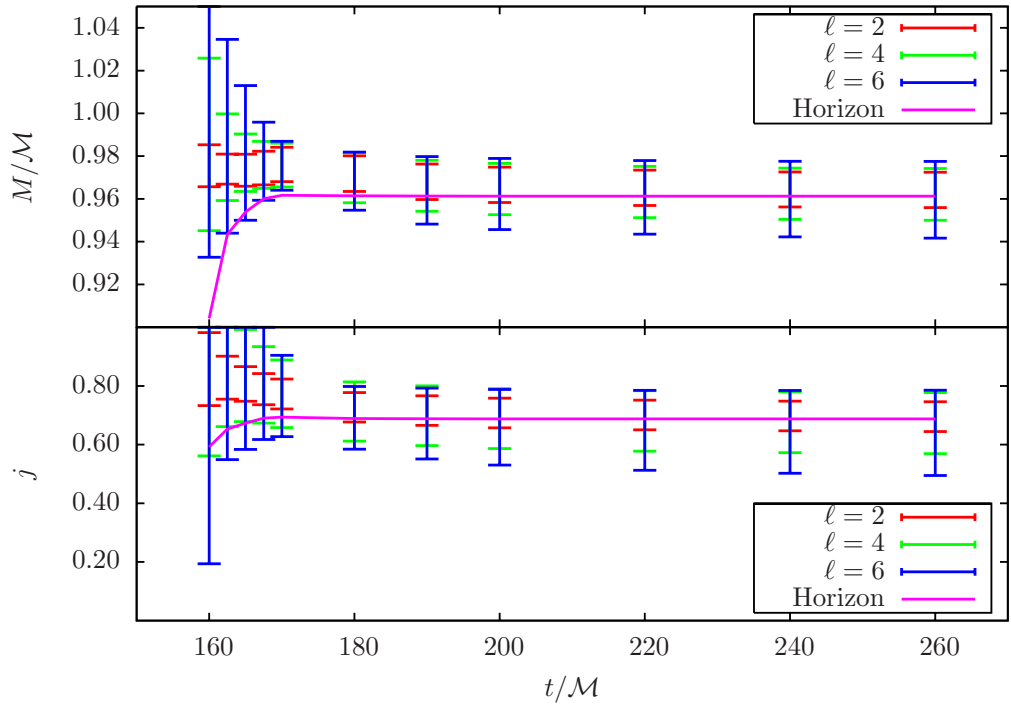}
\caption{The mass and spin parameters extracted from the fundamental mode frequency, for the three 
angular modes. The error bars for the horizon mass and spin is included in the curve 
width. At early times, different choices of $\tau_0$ and $\tau_f$ lead to significant 
modifications in the behavior of $M$ and $j$ as a function of $t$. The data error bars 
in this regime should therefore be considered as an indication of the mass and spin
range on each hypersurface.\label{fig:mj}}
\end{figure*}

Figure \ref{fig:mj} shows the \bh{} mass $M$ and spin $j$ as obtained
from the scalar field evolved on different hypersurfaces labeled by
$t$, for the $\ell=\{2,4,6\}$ modes.
For reference, we also show
the values obtained directly from the dynamical horizon finder.
Both $M$ and $j$ show qualitatively similar behavior, with an initial
transition regime immediately after the first common apparent horizon formation (at
$t=160\mathcal{M}$). About $20\mathcal{M}$ after the common apparent horizon is found
we can reliably extract the spin and mass of the final \bh{} from the
scattered scalar field. 

We want to emphasize
that, while the scalar field modes on the different hypersurfaces
appear almost identical, the extracted mass and spins show much larger
differences due to the strong dependence of $M,j$ on the extracted
quasinormal frequency $\Omega_{\ell m}$
(notice that the error bars in figure \ref{fig:mj} represent not only the 
uncertainty associated with the best-fit procedure, but  also 
the contribution to the error due to finite numerical resolution, to angular mode
extraction at a finite distance from the black hole and to the choice to neglect any
overtones or different angular modes in the frequency extraction;
these contributions will be discussed in section \ref{sec:errors}).

In the transition regime between $160 \mathcal{M}$ and $180 \mathcal{M}$ the fit 
errors (in particular, those from varying the choice of $\tau_0$ and $\tau_f$) become
so large that an accurate extraction of the \bh{} parameters is not
possible anymore. Note that the extraction of $M$ and $j$ at early
times is so sensitive to $\tau_0$ and $\tau_f$ that the associated uncertainty 
dominates the bars shown in figure \ref{fig:mj} for $t \lesssim 180 \mathcal{M}$. 

We conclude that our scattering procedure constitutes an accurate probe of the 
final black hole potential (defined in the perturbative sense as summarized in 
section \ref{sec:scalar}) at times $t \sim 180 \mathcal{M}$, consistent with the results
from the isolated horizon parameters.
Between $160\mathcal{M}$ and $180\mathcal{M}$, 
the \bh{} parameters $M$ and $j$ cannot be reliably extracted; however, the
scalar wave is still showing quasinormal ringing.  In this regime, 
due to the level of uncertainty associated
with our numerical resolution and frequency extraction routine,
we can only provide a broad constraint for the extracted parameters.

\subsection{Effect of finite resolution, extraction radius and number of fitting modes}
\label{sec:errors}
At this point we would like to discuss the different error sources for
the final extraction of the \bh{} parameters $M$ and $j$ from the scalar
field waveforms. We discuss how sensitive the extraction of these
parameters is to the fit range $[\tau_0, \tau_f]$, 
to resolution, to wave extraction radius and to fitting the waveform
to a superposition of more than one quasinormal mode. It is important
to keep in mind that the \bh{} mass and spin parameters show a strong dependence
on the scalar quasinormal frequencies $\Omega$ and hence a small fractional error
in $\Omega$ translates into a much amplified fractional error in $M$ and,
especially, $j$.

\begin{figure}
\vbox{
\includegraphics[width=0.5\textwidth]{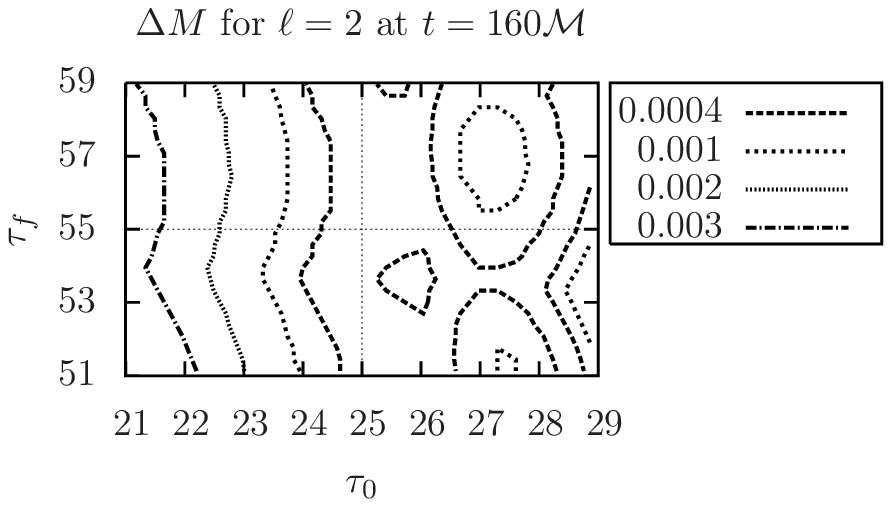}
\includegraphics[width=0.5\textwidth]{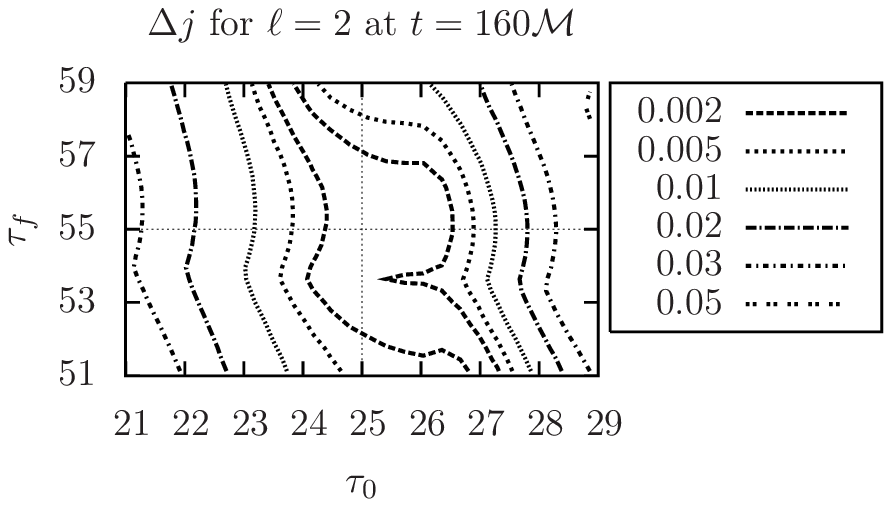}
}
\caption{Contours of the variation of the mass estimate 
$\Delta M(\tau_0,\tau_f)=|M(\tau_0,\tau_f)-M(25\mathcal{M},55\mathcal{M})|$ 
and spin estimate $\Delta j(\tau_0,\tau_f)=|j(\tau_0,\tau_f)
-j(25\mathcal{M},55\mathcal{M})|$
for $\ell=2$ at $t=160\mathcal{M}$, as a function of $\tau_0$ and 
$\tau_f$. The other cases follow similar patterns, with error ranges
improving at later times and degrading for higher 
$\ell$-modes.\label{fig:2d}}
\end{figure}

In order to provide a description of the {\em fitting error},
especially due to the choice of the fitting window $[\tau_0, \tau_f]$,
we select $\tau_0=25\mathcal{M}$ and $\tau_f=55\mathcal{M}$ as
reference values and vary these two
parameters over a $10\mathcal{M}$-wide interval centered around each.
The induced variation in $M$ and $j$ is illustrated in figure
\ref{fig:2d} for a few representative cases: this estimate provides a
quantitative range for the uncertainty in the fitting parameters.  
In the following, we limit ourselves to this pair of reference values and quote the
error associated with this choice.

For the three different {\em resolutions} listed above ($h=\mathcal{M}/44.8$,
$\mathcal{M}/51.2$ and $\mathcal{M}/57.6$),
we find that both the real and the imaginary part of the extracted frequencies 
exhibit a monotonic trend as $h$ decreases, with the difference between neighboring 
resolutions decreasing as $h \to 0$ and the difference between the highest and the 
lowest resolution remaining always under $0.0003\mathcal{M}^{-1}$. 
As for the {\em extraction radius}, 
the six different choices ($\rho=10\mathcal{M}, 20\mathcal{M}, \dots , 50\mathcal{M}$
in addition to the initial $\rho=5\mathcal{M}$) lead to frequency shifts of
the order of $0.005\mathcal{M}^{-1}$. The uncertainty is also smaller if 
the extraction is performed on detectors that are farther from outer and refinement 
boundaries, such as the ones at $5\mathcal{M}$, $10\mathcal{M}$ and $30\mathcal{M}$ 
(the different spatial resolutions characteristic of different extraction radii are not 
dramatically relevant, indicating that the modes are well resolved in all the cases
considered).

Finally, the actual eigenfunctions of the angular part of the wave operator 
in equation (\ref{eq:KG}) on a Kerr spacetime are the (spin-zero) spheroidal harmonics 
$S_{\ell m}$, rather than the spherical harmonics $Y_{\ell m}$ used here; however, as 
discussed in~\cite{1973ApJ}, 
$Y_{\ell m}=S_{\ell m}+(a^2|\Omega|^2) \sum_{\ell^\prime \neq l} B_{\ell^\prime m} S_{\ell^\prime m}+O(a^4|\Omega|^4)$, 
so that the use of spherical harmonics only causes a modest amount of 
mode mixing in the extracted waveforms (additionally, Berti et al.~\cite{Berti:2005gp} find
that this expansion is surprisingly accurate out to values of $a|\Omega|$ close to unity).
To determine the bias caused by such mode mixing on the best-fit value of the 
fundamental frequency,
we extend the fitting function (\ref{eq:fitfunc}) to a superposition of two and three
different angular modes.
The change in the complex frequency of the fundamental mode due to the inclusion of a different number of 
modes is under $0.0006 \mathcal{M}^{-1}$.
Unlike the error due to the frequency fitting, these last three sources of error have virtually 
no fluctuations from hypersurface to hypersurface.
For all times,
the cumulative errors associated with the determination of the complex frequency
(including the last three sources of error but not the uncertainty 
introduced by the fitting procedure)
then amount to $\sim 0.005 \mathcal{M}^{-1}$. This number is propagated to $M$ and $j$ via
$\Delta M = M(\Omega+\Delta \Omega)-M(\Omega)$ and $\Delta j = j(\Omega+\Delta \Omega)-j(\Omega)$
which yields:
\begin{center}
\begin{tabular}{lllll}
$\Delta M=0.02 \mathcal{M}$  &,\quad& $\Delta j=0.1  $   &\;& for $\ell=2$ \\
$\Delta M=0.02 \mathcal{M}$  &,\quad& $\Delta j=0.15 $   &\;& for $\ell=4$ \\
$\Delta M=0.03 \mathcal{M}$  &,\quad& $\Delta j=0.2  $   &\;& for $\ell=6$ \,.
\end{tabular}
\end{center}
Finally, these numbers are summed in quadrature to the errors in $M$
and $j$ due to the fitting procedure (shown in figure \ref{fig:2d}), 
yielding the error bars in figure \ref{fig:mj}.


\section{Discussion and conclusions}\label{Conclusions}
In the previous section, we have presented the result of an experiment involving the
scattering of a massless scalar field off the curvature of a spacetime containing two 
coalescing black holes. This spacetime was generated numerically by a full 3D black hole
simulation following the moving puncture paradigm. Each spatial hypersurface 
obtained served as a fixed background for the propagation of the scalar field 
as described in section \ref{sec:sim} and
the behavior of the scalar field was analyzed in each case.

We observed the emergence of quasinormal ringing on each spatial slice
containing a common apparent horizon regardless of whether the merged
system had already settled down to a single Kerr black hole. While we
could not reliably extract the final \bh{} parameters immediately after
the formation of a common apparent horizon ($t=160\mathcal{M}$), we
could still clearly identify quasinormal ringing even in this early
phase where the merged \bh{} is strongly excited and the scattering
potential deviates from that of Kerr. This finding agrees with
earlier results where scalar perturbations on Vaidya spacetimes were
studied and quasinormal ringing could be identified even on
time-dependent black hole backgrounds~\cite{Shao:2004ws}.
Our results and interpretation are also consistent with the evidence 
from the Close Limit approximation~\cite{Price:1994pm}, where a pair of black holes in 
a head-on collision was shown to have already entered the perturbative regime when a common apparent 
horizon surrounded them. 
Based on this observation, the scattering of a test scalar field
from the merged \bh{} could provide information about the final \bh{} state 
before the spacetime has settled down to its final state.

Soon after horizon formation (i.e. for times $t \gtrsim
180\mathcal{M}$) we were able to obtain estimates for the mass and
spin of the final \bh{} from the scalar field probe in good agreement
with direct isolated horizon measures. This indicates that, at this
time, the dynamics surrounding the \bh{} has settled down sufficiently
that the scattering potential for the scalar field is essentially that
of the final \bh{}, even though there is still non-trivial dynamics in the
vicinity of the horizon.

A more exhaustive analysis of the early-time scalar quasinormal
ringing waveforms before $t=180 \mathcal M$ is necessary to obtain more conclusive results in
the non-linear, merger phase, and is left for future work.

\acknowledgments
The authors wish to thank Emanuele Berti for providing the scalar 
quasinormal frequencies used in this work and for his comments on the 
manuscript, Erik Schnetter 
for the \texttt{ReflectionSymmetry} and \texttt{IsolatedHorizon} modules, and 
Marcus Ansorg for the initial data generator \texttt{TwoPunctures}.
They also acknowledge the support of the Center for
Gravitational Wave Physics funded by the National Science Foundation
under Cooperative Agreement PHY-0114375. 
This work was partially supported by NSF grant
PHY-0555436 and PHY-0653443.
The simulations presented in this paper were carried out under allocation 
TG-PHY060013N at NCSA.

\bibliography{biblio}


\end{document}